
\input phyzzx

\nopagenumbers
\tolerance=10000
\Pubnum={YNUE/PH/92/01}
\date= {}
\titlepage
\title{\bf QUANTUM HAMILTONIAN REDUCTION OF SUPER KAC-MOODY ALGEBRA II  }
\author{Tetsuji Kuramoto}

\address{ Department of Physics,
\break
Faculty of Engineering, Yokohama National University,
\break
 156 Tokiwadai Hodogaya, Yokohama, Japan}
\abstract
\indent The quantum Hamiltonian reduction on the OSp(1,2) super
Kac-Moody algebra is described in the BRST formalism.
 Using a free field representation of the KM currents,
the super Kac-Moody algebra is shown to be reduced to a superconformal one
via the Hamiltonian reduction.
 This reduction is manifestly supersymmetric because of supersymmetric
constraints imposed on the algebra.

\endpage

\def\ss#1{\scriptscriptstyle #1}
\def\ssr#1{\scriptscriptstyle {\rm #1}}
\def\alphap{\alpha_{\scriptscriptstyle +}}
\def\wt#1{{\widetilde #1}}
\def\wh#1{{\widehat #1}}

\def\qb{Q_{\ssr B}}
\def\hal{{1\over2}}
\def\ts#1{{\textstyle #1}}
\def\psip{\psi_{\ssr {phys}}}
\def\hp{{\cal H}_{\ssr {phys}}}
\def\bra#1{\vert\, #1 >}
\def\sk{\sqrt{2{\wh k}}\>}

\baselineskip=18pt
\REF\BEL{A.A. Belavin, in: Quantum String Theory, Springer Proc. in Physics
Vol 31, eds. N. Kawamoto and T. Kugo (Springer-Verlag, Berlin, Heidelberg,
1988 ) p132.}
\REF\BOC{M.B. Bershadsky and H. Ooguri, Commun. Math. Phys. 26 (1989) 49.}
\REF\POLY{A.M. Polyakov, Mod. Phys. Lett. A2 (1987) 893.   }
\REF\BOS{M.B. Bershadsky and H. Ooguri, Phys. Lett. 229B (1989) 374.   }
\REF\KK{K. Kimura, preprint RIMS-811 (1991). }
\REF\TK{T. Kuramoto, preprint QMW/PH/90/20 ;  in: Proc. of KEK workshop
" Superstrings and Conformal Field Theory " eds. M. Kobayashi
and M. Kato  (KEK, July 1991) p18.}
\REF\GERA{A. Gerasimov, A. Marshakov and A. Morozov, Phys. Lett. 236B (1990)
269.    }

\REF\SKM{V.G. Kac and T. Todorov, Commun. Math. Phys. 102 (1989) 337
        .   }
\REF\KOS{T. Kugo and I. Ojima, Suppl. Prog. Theor. Phys. 66 (1979) 1.}
\REF\TKS{T. Kuramoto, Nucl. Phys. B346 (1990) 527.}
\REF\SCF{ M.A. Bershadsky, V.G. Knizhnik and M.G. Teitelman, Phys. Lett.
            151B (1985) 31 ;\hfill\break
         G. Mussardo, G. Sotkov and M. Stanishov, Phys. Lett. 195B (1987) 397.}
\REF\SUPER{H. Eichenherr, Phys. Lett. B151 (1985) 26;\hfill\break
            D. Friedan, Z. Qui and S. Shenker, Phys. Lett. B151 (1985) 37. }
\REF\BRST{M. Kato and K. Ogawa, Nucl. Phys. B212 (1983) 443;\hfill\break
         M. Ito, T. Morozumi, S. Nojiri and S. Uehara, Prog. Theor. Phys.
          75 (1986) 934;\hfill\break
         K. Itoh, Nucl. Phys. B342 (1990) 449.}
\REF\F{G. Felder, Nucl. Phys. B317 (1989) 215.}
\REF\KS{ Y. Kazama and S. Suzuki, Nucl. Phys. B321 (1989) 232;\hfill\break
        C.M. Hull and B. Spence, Phys. Lett. B241 (1990) 357.}

\footline={\hss\tenrm\folio\hss}
\pageno=1

\chapter {Introduction }

  One of interesting properties of  conformal field theory has been
observed by applying the Hamiltonian reduction to a SL(2,R) Kac-Moody
algebra  [\BEL,\BOC]; an irreducible representation of the conformal algebra is
obtained from one of the SL(2,R) KM algebra by imposing a certain
constraint.  In this constraint system, which can be
described in the BRST formalism like gauge field theories,
the KMA algebra is not observable and so is hidden behind a physical
symmetry of the conformal algebra. Thus one calls the SL(2,R) KMA
a hidden symmetry of  conformal field theory. This discovered connection
between the KM and conformal algebras explains the appearance of the
SL(2,R) KM algebra in the effective action of  2-d gravity in the
light-cone gauge [\POLY].

  Seeking the hidden symmetry of  superconformal field theory, the
Hamiltonian reduction is applied to an  OSp(1,2) KM algebra [\BOS].
The consistency of  imposed constraints forces to introduce
a Majorana fermion in addition to the KM currents because the constraints
satisfy a nontrivial algebra unlike in the SL(2,R) case.
One then succeeds in reducing the combined system of the OSp(1,2) KMA
and the fermion to a superconformal field theory.
Although that system is the smallest one that is reduced to the
superconformal field theory, there are the following unsatisfactory
points involved ; (i) the fomulation is not manifestly supersymmetric
and so the relation of the extracted superconformal algebra with the
original one  is obscure. (ii) Since the fermion has
been added to the space of the OSp(1,2) KMA, it is questionable to
conclude that the hidden symmetry of superconformal field theory is
the OSp(1,2) KMA.  We would like to consider that the addition of
the fermion into the system suggests  a larger hidden symmetry than
the OSp(1,2) KMA.
In the present paper, thus, we give a supersymmetric formulation
of the  quantum Hamiltonian reduction on the OSp(1,2) super KMA.
( See [\KK] for the classical one.)

   Using a free-field realization of the OSp(1,2) KMA and explicitly
solving the constraints on the super KM (SKM) currents, we tried in [\TK]
to show that two unconstrainted SKM currents are directly reduced
into the supercurrent and stress-energy tensor of superconformal field theory.
Although this approach is successful in the SL(2,R) case [\GERA], we did not
reach any definite results because of the subtlety of  the normal
ordering of higher-order terms in the SKM currents
when the solved constraints are substituted into them.
To avoid this problem, we  apply the BRST formalism to our system
of the super KMA with constraints on superspace currents.

   This paper is organized as follows. Section 2 reviews the main results of
super Kac-Moody algebra. In Section 3 we propose constraints on superspace
SKM currents, which allow the OSp(1,2) super KMA to be supersymmetrically
reduced into a superconformal algebra, introducing a c-number function
 $\zeta(z)$. The function $\zeta(z)$ plays the same important role
as the constant $1$ in the constraint $J^{(-)}(z)-1=0$ of the SL(2,R) case.
After the introduction of pairs of ghosts and antighosts corresponding to
the constraints, a BRST charge of our system is constructed.
In Section 4 we show that the proposed constraints really reduce a space
of the super KMA into a superconformal field theory, using the technique
developed for the proof of the no-ghost theorem in string theory.
The same constraints as in the OSp(1,2) case are obtained from our
constraints on SKM currents by substitutting the fermion constraints into
them. In Section 5 we exhibit that a twisted superconformal tensor of
the constraint system really becomes in the physical subspace one of
superconformal field theory.  When the function $\zeta(z)$ depends on $z$,
the BRST invariance requires some operator to be added to the twisted
superconformal tensor. We discuss a physical meaning of the added term.
Section 8 contains a brief conclusion and some comments.

\chapter {OSp(1,2) Super Kac-Moody Algebra}

 Let us start with the OSp(1,2) Kac-Moody algebra with central charge $k$.
This KM algebra has three bosonic currents $ J^{(a)}(z)$ for $a=0,\pm 1$,
which generate the SL(2,R) KM subalgebra, and two fermionic currents
$J^{(\alpha)}(z)$ for $\alpha =\pm {1\over 2}$. The KM currents
$J^{({\ss A})}$  for $A=0,\pm 1,\pm {1\over2}$ satisfy
the following operator product expansion ( OPE )
 $$ J^{({\ss A})}(z)J^{({\ss B})}(w)\sim {\eta ^{\ss {AB}}{ k\over 2}
                                            \over (z-w)^2 }
                   +{f^{\ss {AB}}{}_{\ss C} J^{({\ss C})}(w)\over z-w}
                                         \eqn \jope$$
where $\eta^{\ss {AB}}$ and $f^{\ss {AB}}{}_{\ss C}$ are the metric and
structure constant of the algebra.

  A KM algebra can be supersymmetrized by introducing fermions in the adjoint
representation which become supersymmetric partners of super KM currents
[\SKM].
In the present case we then introduce three anticommutting fermions
$\psi ^{(a)}$ for $a =0,\pm 1$ and two commutting fermions $\psi ^{(\alpha)}$
for
$\alpha =\pm{1\over2}$, corresponding to the bosonic and fermionic currents
$J^{(a)}$ and $J^{(\alpha)}$ respectively. The superspace currents
${\bf J}^{({\ss A})}$ are constructed as
 $$ {\bf J}^{({\ss A})}(z,\theta)=\psi ^{({\ss A})}(z)
                                +\theta {\wh J}^{({\ss A})}(z)
                                                      \eqn \superc $$
and generate the super KM algebra (SKMA) with
central charge ${\wh k}$, satisfying  the following OPE
$$ {\bf J}^{({\ss A})}(z_1,\theta_1) {\bf J}^{({\ss B})}(z_2,\theta_2)\sim
                         {{{\wh k}\over 2}\eta^{\ss AB}\over z_{12}}
       +{\theta_{12}f^{\ss AB}{}_{\ss C}{\bf J}^{({\ss C})}(z_2, \theta_2)
                         \over z_{12}}
                                                           \eqn \opesj$$
where $z_{12}\equiv z_1-z_2-\theta_1\theta_2$ and
          $\theta_{12}\equiv\theta_1-\theta_2$.
{}From its operator product with the fermion in the OPE \opesj, one finds that
the super KM (SKM) current $ {\wh J}^{({\ss A})}$ consists of two independent
parts:
 a current part $J^{({\ss A})}$ with central charge $k$ and a fermion part
$J^{({\ss A})}_f$ with central charge $3\over2$
$$ {\wh J}^{({\ss A})}(z)=J^{({\ss A})}(z)+J^{({\ss A})}_f(z).\eqn \totalj$$
The fermion current $J^{({\ss A})}_f$ is given by
       $$ J^{({\ss A})}_f={\textstyle {1\over {\wh k}}}f^{\ss A}{}_{\ss BC}
                           \,\colon\psi ^{({\ss C})}\psi ^{({\ss B})}\colon .
                                                             \eqn \fermij $$
Then the total central charge  ${\wh k}$ of the SKMA is equal to a sum of
those central charges  $ k+{3\over2}$.

One can construct an N=1 superconformal algebra (SCA) from  superspace KM
currents. The super stress-energy tensor ${\bf T}^{\ssr SKM}$ is given by
$$ {\bf T}^{\ssr SKM}(z, \theta)={\textstyle {1\over2}}G^{\ss SKM}(z)
                          +\theta T^{\ssr SKM}(z) \eqn \st $$
where the supercurrent and stress-energy tensor are written in the form
  $$\eqalign{G^{\ssr SKM}(z)&={\textstyle{2\over{\wh k}}}
              \bigl( \eta _{\ss AB}J^{({\ss B})}\psi^{({\ss A})}
                   -{\textstyle {1\over 3{\wh k}}} f_{\ss ABC}
                   \>\colon\psi ^{({\ss A})}\psi^{({\ss B})}
                       \psi^{({\ss C})}\colon \bigr) \cr
    T^{\ssr SKM}(z)&={\textstyle {1\over{\wh k}}}
                     \bigl(\eta_{\ss AB}\>\colon J^{({\ss B})}J^{({\ss A})}
                   \colon -\eta_{\ss AB}\> \colon \psi^{({\ss B})}
                 \partial\psi^{({\ss A})}\colon \bigr)
                             \cr} \eqn \sts $$
Since the metric and structure constant of the OSp(1,2) contain the unfamiliar
 antisymmetric and symmetric parts  caused by fermionic elements, we carefully
checked the ordering of operators in the expressions \fermij\ and \sts.
The conformal anomaly $c_{\ssr SKM}$ of the SCA is easily obtained as
$$ c_{\ssr SKM}={1\over2}+{k\over k+{3\over2}} \eqn \skma $$
by calculating the OPE between two supercurrents
 $$ G^{\ssr SKM}(z)G^{\ss SKM}(w)\sim {{2\over3}\>c_{\ssr SKM}\over (z-w)^3}
                                   +{2\>T^{\ssr SKM}(w)\over z-w}
   \eqn \ggope $$
With respect to the stress-energy tensor $T^{\ssr SKM}$ of Sugawara form, all
the superspace currents \superc\  have the same dimensions $1\over2$;
$\psi ^{({\ss A})}$ of dimension $1\over2$ and ${\wh J}^{({\ss A})}$
of dimension $1$.

The Hilbert space of a SCA is classified into two sectors, the Ramond sector
with expansion modes of $G(z)$ integers ($G(z)=\sum G_n z^{-n-{3\over2}}$)
and the Neveu-Schwarz sector with expansion modes half-integers
($G(z)=\sum G_{n+{1\over2}}z^{-n-2}$). In the SKMA, along with the integer mode
expansion of the currents ${\wh J}^{({\ss A})}$, the fermions
$\psi^{({\ss A})}$ are correspondingly expanded as
$$ \psi^{({\ss A})}(z)=\sum \psi ^{({\ss A})}_{n+\kappa}
z^{-n-\kappa-{1\over2}}
                              \eqn \expf $$
with $\kappa=0$ for the Ramond sector and $\kappa={1\over2}$ for the
Neveu-Schwarz sector.

\chapter  {Quantum Hamiltonian Reduction on Super Kac-Moody Algebra }

   In the Hamitonian reduction of the SL(2,R) and OSp(1,2) KMA's,
the stress-energy tensor of Sugawara form was modified by the addition of
a current linear term [\BOC,\BOS].  For the case of the SKMA, an appropriate
super
stress-energy tensor can be constructed  by twisting one of Sugawara form
 ${\bf T}^{\ssr SKM}(z,\theta)$ with a superspace current
${\bf J}^{(0)}(z, \theta)$ as
$$ {\bf T}(z, \theta)\equiv {\bf T}^{\ssr SKM}-\partial {\bf J}^{(0)}.
\eqn \tskm $$
The twisted super stress-energy tensor ${\bf T}$ satisfies a SCA with
conformal anomaly $c=c_{\ssr SKM}-6{\wh k}$. With respect to ${\bf T}$,
the superspace current ${\bf J}^{({\ss A})}$ is of dimension ${1\over2}+A$;
$\psi^{({\ss A})}$ of dimension ${1\over2}+A$ and ${\wh J}^{({\ss A})}$ of
dimension $1+A$. The significant effect of the shift of the conformal
dimension will be discussed later.

  In order to extract a SCA in a supersymmetric way, we put the following
constraints on superspace currents,
$$\eqalign{{\bf J}^{(-)}(z, \theta)&=0 \cr
         {\bf J}^{(-{1\over2})}(z, \theta)&-{\ts \sk}\zeta(z) =0 \cr } \eqn
\sjc $$
which are equivalent to
$$\eqalign{\psi^{(-)}(z)&=0 \cr \psi^{(-{1\over2})}(z)&-{\ts \sk}
            \zeta(z) =0 \cr}\qquad
\eqalign{{\wh J}^{(-)}(z)&=0\cr  {\wh J}^{(-{1\over2})}(z)&=0 \cr} \eqn \fajc$$
Here we have introduced the arbitrary c-number function $\zeta (z)$ of $z$.
The function $\zeta$ is required by  the first constraint in the second
column of \fajc\ to be of the same properties as the fermion
$\psi^{(-{1\over2})}$, in particular, commutting with $\psi^{(\pm{1\over2})}$
and anticommutting with $\psi^{(a)}$ for $a=0,\pm$. Those superspace
constraints \sjc\ are consistent and closed because the superspace current
${\bf J}^{(-{1\over2})}$ is of dimension $0$ and only their nontrivial
operator product is
$$ \bigl(\>{\bf J}^{(-{1\over2})}(z_1, \theta_1)-{\ts \sk}\zeta\>\bigr)\>
  \bigl(\> {\bf J}^{(-{1\over2})}(z_2, \theta_2)-{\ts \sk}\zeta\>\bigr)\sim
 {-2\over z_{12}}\>\theta_{12}{\bf J}^{(-)}(z_2, \theta _2).
  \eqn\sca $$
This operator product also indicates that one can not consistently fix
the current ${\wh J}^{(-)}$ of dimension $0$ to be any nonvanishing constant
like in the SL(2,R) case.

  We introduce two superspace ghost systems,
$({\bf b}(z, \theta), {\bf c}(z, \theta))$ of dimensions $(-{1\over2}, 0)$
and $( \beta (z, \theta), \gamma (z, \theta))$ of dimensions
$(0, {1\over2})$, corresponding to the superspace constraints \sjc,
to complete the standard quantization process of  constraint system.
The superspace ghosts are expanded in  $\theta$ as
$$ \eqalign{{\bf b}(z, \theta)&=\beta ^++\theta b \cr
                \beta (z, \theta)&={\wt b}+\theta \beta_+ \cr}\qquad
\eqalign{{\bf c}(z, \theta)&=c_z+\theta\gamma_{z+} \cr
                \gamma (z, \theta)&=\gamma_++\theta{\wt c}_z \cr}
\eqn\ghost $$
The pairs of ghosts, $(b, c_z)$  and $({\wt b}, {\wt c}_z)$
both of dimensions $(0, 1)$, are anticommutting and the pairs,
$(\beta ^+, \gamma_{z+})$ of $(-{1\over2}, {3\over2})$ and
$(\beta_+, \gamma_+)$ of $({1\over2},\hal)$, are commutting.
Their total super stress-energy tensor with conformal anomaly
$c_{\scriptscriptstyle {\rm ghost}}=6$ is written in a supersymmetric form
 $${\bf T}^{\ssr {ghost}}(z, \theta)=\bigl({\bf c}\>D^2{\bf b}
-{\textstyle {1\over2}}D{\bf c}\>\cdotp D{\bf b}+{\textstyle {1\over2}}D^2{\bf
c}
\>\cdotp {\bf b}\bigr)
+{\textstyle {1\over2}}\bigl(\gamma \>D^2 \beta -D \gamma \>\cdotp
                                      D \beta \bigr)
\eqn \ghostst$$
where $D\equiv {\partial\over\partial \theta}+\theta\partial$ is a covariant
derivative in the superspace.

  Combining the superspace constraints \sjc\ with the superspace
ghost systems \ghost\ and using its closed subalgebra \sca, a BRST charge
of our system is constructed in the standard manner,
$$ Q_{\ssr B}=\oint\> dzd\theta\> \bigl\{\> {\bf c}\>{\bf J}^{(-)}
      +\gamma\> ({\bf J}^{(-{1\over2})}-{\ts \sk}\zeta )-\gamma^2\>{\bf
b}\bigr\}
\eqn\qbs $$
which is  written after the integration of $\theta$ as
$$  Q_{\ssr B}=\oint \> dz \bigl\{ c_z\,{\wh J}^{(-)}
                      +\gamma_+\,{\wh J}^{(-{1\over2})}
       +\gamma_{z+}\,\psi^{(-)}+{\wt c}_z\,(\psi^{(-{1\over2})}-{\ts \sk}
      \zeta)
      -\gamma^2_+\,b-2\gamma_+\,{\wt c}_z\,\beta^+\bigr\}. \eqn\qbc $$
One can easily verify the nilpotency of the BRST charge $Q_{\ssr B}$.
The BRST charge $Q_{\ssr B}$ plays a central role in the BRST formalism,
giving a criterion for a physical operator ${\cal O}_{\ssr {phys}}$ as
$[\,Q_{\ssr B},{\cal O}_{\ssr {phys}}\,]_{\pm}=0$.
Since $Q_{\ssr B}$ sums over all generators of gauge symmetry multiplied by
the corresponding antighosts, this  implies that the physical operator
does not  contain any gauge freedoms.
Furthermore, the physical operator can be written in the
form of the summation over a genuine  and a BRST trivial parts as
$${\cal O}_{\ssr {phys}}= {\cal O}_{\ssr G} + [\,Q_{\ssr B}, \ast\,]_{\pm}
 \eqn\physo$$
where ${\cal O}_{\ssr G}$ can not be expressed in the form of the
(anti)commutation with $\qb$ and $\ast$ indicates some operator [\KOS].
Also, a physical subspace $\hp$ is defined as a quotient space
$$\hp\equiv\{\;\bra{\upsilon}\>;\; \qb\bra{\upsilon}=0\,\}\Big/
\{\;\bra{\star}\>;\; <\star\,\bra{\star}=0\;
\>  \hbox{and}\;\>\qb\bra{\star}=0\, \}. \eqn\dphys$$
We will find out the physical symmetry algebra which is extracted
from the SKMA through the above Hamiltonian reduction.

Let us first mention the fate of the SKMA.  Of the SKM currents,
only ${\bf J}^{(-)}$ commutes with the BRST charge $\qb$ but it has no
genuine part because ${\bf J}^{(-)}(z, \theta)=\{\qb,{\bf b}(z, \theta)\}$.
Thus the SKMA is not any symmetry algebra in the physical
subspace. Since the  ghost systems have been introduced into the Hilbert
space of the SKMA, a total super stree-energy tensor ${\bf T}^{\ssr {total}}$
of the whole system is a sum of ones for the twisted SKMA and ghosts,
\tskm\ and \ghostst,
$$ {\bf T}^{\ssr {total}}(z, \theta)=({\bf T}^{\ssr {SKM}}
                         -\partial{\bf J}^{(-)})
                                 +{\bf T}^{\ssr {ghost}} \eqn\totalt $$
with conformal anomaly $c_{\ssr {total}}=(c_{\ssr {SKM}}-6{\wh k})
+c_{\ssr {ghost}}$.  Some algebraic calculation shows that
 $$  \{\>\qb,{\bf T}^{\ssr {total}}(z, \theta)\>\}=0. \eqn\qbt $$
The total super stress-energy tensor ${\bf T}^{\ssr {total}}$ also contains
a nonvanishing genuine part, as we will show later. Thus the superconformal
algebra generated by ${\bf T}^{\ssr {total}}(z, \theta)$  is the  extracted
symmetry of the physical subspace ${\cal H}_{\ssr{phys}}$  from the SKMA.

  The currents and fermions ${\wh J}^{({\ss A})}$ and $\psi^{({\ss A})}$
of twisted dimensions $1+A$ and ${1\over2}+A$ are expanded as
$$  \psi^{({\ss A})}(z)=\sum ^{\infty}_{n=-\infty}
           \psi^{({\ss A})}_{n+\kappa}z^{-n-\kappa-{1\over2}-A},
   \qquad\qquad {\wh J}^{({\ss A})}(z)=\sum^{\infty}_{n=-\infty}
          {\wh J}^{({\ss A})}_n z^{-n-1-A}. \eqn\expa $$
In the expansion \expa,  the bosonic currents are even under
the transformation $z\rightarrow e^{2\pi i}z$  but the fermionic odd.
This indicates that the expansion \expa\ breaks the OSp(1,2) (global)
symmetry.
On the other hand,  to keep the symmetry,  we may use the other expansion of
the currents and fermions, in which the fermionic currents are twisted as
   ${\wh J}^{(\alpha)}(z)=\sum {\wh J}^{(\alpha)}_{n+{1\over2}} z^{-n-{1\over2}
-1-\alpha}$ and the corresponding fermions are expanded as
$\psi^{(\alpha)}(z)=\sum \psi^{(\alpha)}_{n+{\bar \kappa}}z^{-n-{\bar
\kappa}-{1\over2}-\alpha}$  where ${\bar \kappa}={1\over2}$ for the Ramond
sector and ${\bar \kappa}=0$ for the Neveu-Shwarz sector [\TKS].
In the present paper  the former expansion \expa\ will be used.
The constraint $\psi^{(-\hal)}(z)-\sk\zeta=0$, thus, requires the function
$\zeta$ of $z$
to have integer  power in  $z$ for the Ramond sector and half-integer for
the Neveu-Schwarz sector.  For simplicity  the function $\zeta$ will be
chosen to be constant $\zeta_0$ and $\zeta_{\hal}z^{-\hal}$
for the Ramond and Neveu-Schwarz sectors respectively. The ghost systems
$(\beta^+, \gamma_{z+})$ and $({\wt b},{\wt c}_z)$ corresponding to the
fermion constraints in \fajc\ are expanded as
$$\eqalign{\beta^+(z)&=\sum
^{\infty}_{n=-\infty}\beta^+_{n+\kappa}\>z^{-n-\kappa
+\hal}\cr
{\wt b}(z)&=\sum ^\infty_{n=-\infty}{\wt b}_{n+\kappa}\>
   z^{-n-\kappa} \cr}\qquad\qquad
\eqalign{\gamma_{z+}(z)&=\sum ^\infty_{n=-\infty}\gamma_{z+}\,{}_{n+\kappa}\>
z^{-n-\kappa-{3\over2}}\cr
 {\wt c}_z(z)&=\sum ^\infty_{n=-\infty}{\wt c}_z\,{}_{n+\kappa}\>
      z^{-n-\kappa-1}.
 \cr} \eqn\ghostexp $$

The current constraints in \fajc\ can be simplied by using the fermion ones as
$$\eqalign{0&={\wh J}^{(-)}=J^{(-)}+\hal \zeta ^2\cr
   0&={\wh J}^{(-\hal)}=J^{(-\hal)}-{\textstyle{\sqrt{2\over {\wh k}}}}
             \psi^{(0)}\zeta \cr} \eqn\rcc $$
The reduced constraints \rcc\ appeared in [\BOS], which showed
that a Hilbert space of the OSp(1,2) KMA and a fermion $\psi$ are reduced
to that of SCA by imposing the constraints \rcc.
Since  $\psi$ can be identified as  $\psi^{(0)}\zeta$
in the constraints \rcc, the Hilbert space of the KMA  and fermion is
obviously only a subspace of our Hilbert space of the SKMA.

\chapter {BRST Analysis on Hilbert  Space of SKMA and Ghost Systems}

 The following calculation on the total conformal anomaly
$c_{\ssr {total}}$ suggests a connection of the physical subspace
with a Hilbert space of the super Coulomb gas [\SCF] which gives a free field
realization of superconformal models  beacuse a conformal anomaly
countes  dynamical degrees of freedom in the system.
The expression of the anomaly $c_{\ssr {total}}$ can be rewritten as
$$ \eqalign{c_{\ssr {total}}&=(c_{\ssr {SKM}}-6{\wh k})+c_{\ssr {ghost}}\cr
       &={\textstyle {3\over2}}\bigl\{\> 1-2\,(\sqrt{2k+3}-
                                        {1\over\sqrt{2k+3}})^2\>
               \bigr\}\cr}    \eqn\tca $$
The super Coulomb gas is a system of a scalar field and a Majorana fermion
with  backgroud charge $\alpha_0$,  whose anomaly is
$c^{\ssr {SCFT}}={3\over2}(1-8\alpha_0^2)$.  The expression \tca\ of
$c_{\ssr {total}}$  is obtained from the anomaly $c^{\ssr {SCFT}}$
 by replacing $\alpha_0$ with $\alpha_0(k)
                \equiv\hal (\sqrt{2k+3}-{1\over\sqrt{2k+3}})$.
This implies  that the physical subspace is identified as the super Coulomb
gas with  background charge $\alpha_0(k)$, as we will see in
the following sections.
When $k+1={1\over q+2}$ or $-{1\over q+4}$ is satisfied,  the corresponding
physical subspace gives the free field representaion of an unitary
superconformal minimal model with anomaly $c^{\ssr SCFT}(q)={3\over2}\,
  (1-{8\over (q+2)(q+4)})$ for a positive intger $q$ [\SUPER].

 In the SL(2,R) case, its free field representation is a powerful
tool  to analyze the structure of a Hilbert space of the KMA
with a constraint.
To realize the OSp(1,2) KM algebra in terms of free fields,
one needs a scalar field $\phi$, a pair of bosonic ghosts $(\beta , \gamma )$
of dimensions $(1, 0)$, and a pair of fermionic ghosts $(\psi , \psi^+ )$
of dimensions $(0, 1)$. The OSp(1,2) currents are expressed [\BOS] as
$$\eqalign{ J^{(-)}(z)&=\beta \qquad\qquad\qquad\qquad\qquad\qquad
               J^{(-{1\over 2})}(z)=\beta\psi-\psi^+  \cr
  J^{(0)}(z)&=\beta\gamma-{\textstyle {i\over2}}\alphap\partial\phi
                           +{\textstyle {1\over2}}\psi\psi^+ \cr
   J^{(+)}(z)&=-\beta\gamma^2+i\alphap\gamma\partial\phi-\gamma\psi\psi^+
                  +k\partial\gamma-(k+1)\psi\partial\psi  \cr
   J^{({1\over2})}(z)&=\gamma(\psi^+-\beta\psi)+i\alphap\psi\partial\phi
                        +(2k+1)\partial\psi \cr}\eqn \ffc     $$
where $\alphap =\sqrt{2k+3}$. From now on we use the  free field
expressions \ffc\  as the current part $J^{({\ss A})}$ of the SKM
current  \totalj. The total super stress-energy tensor
${\bf T}^{\ssr {total}}$ then
is rewritten in terms of the free fields only.
The pairs of the ghosts $(\beta, \gamma)$ and $(\psi,\psi^+)$ have
twisted dimensions $(0,1)$ and $(\hal ,\hal)$ with respect to
${\bf T}^{\ssr {total}}$.  From the commutation relation of
the SKM currents with the fermions, the pair of the fermionic ghosts
$(\psi,\psi^+)$ are commutting with $\psi^{(a)}$ for $a=0,\pm$
but anticommutting with $\psi^{(\alpha)}$ for $\alpha=\pm\hal$
and they are untwisted because the fermionic currents $J^{(\alpha)}$ have
integer modes. The Ramond and Neveu-Schwarz sectors are obtained
by untwisting and twisting the superpartners $\psi^{({\ss A})}$ of
the SKM currents as  the first expression in \expa.

  One can readily see that physical fields (anti)commutting with the
BRST charge $\qb$ are a scalar field $\phi(z)$ and a composite fermion
$\psi_{\ssr {phys}}(z)\equiv{i\over\sqrt{2\beta}}\,(\beta\psi+\psi^+)$
whose operator product is $\psi_{\ssr {phys}}(z)\psi_{\ssr {phys}}(w)
\sim {1\over z-w}+o(z-w)$.
The factor $i\over\sqrt{2\beta}$ has been multiplied for the normalization.
 Since the field $\beta$ itself has
a curious property of commutting ghost, one must carefully define the
square root function $\sqrt{\beta}$.  From the first constraint
in \rcc\ and the free field expression of $J^{(-)}$ in \ffc,
one can  derive a relation $\beta=-\hal \zeta^2+$ [a BRST trivial term]
in the physical subspace.
This relation gives the precise definition of the square root function  as
$$ \sqrt{\beta}\equiv{i\over\sqrt{2}}\zeta +[\qb,\ast\>]. \eqn\sqrb $$
Thus the function $\sqrt{\beta}$ is anticommutting with the fermionic ghosts
 $(\psi,\psi^+)$ and the physical fermion $\psip$ with the fermions
$\psi^{({\ss A})}$.

  In the following we will show that the physical subspace $\hp$ is
spanned only by excitations of the scalar field and the fermion, $\phi$
and $\psip$, a space of which is referred as ${\cal H}(\phi,\psip)$.
Obviously, $\qb\, \bra{\Omega} =0$ holds for any state $\bra{\Omega} $ of the
space ${\cal H}(\phi,\psip)$. Then it must be proved that any state
$\vert\Psi >$ annihilated by $\qb$ is written in the form
$$ \bra{\Psi }={\cal P}\bra{\Psi}+\qb \bra{\ast} \eqn\physp$$
where the operator $\cal P$ is a projection onto the space
${\cal H}(\phi,\psip)$ from the whole Hilbert space.
The vacuum $\bra 0$ is defined by
$$\varphi_n\>\bra 0 =0 \qquad {\overline \varphi}_m\bra 0 =0 \eqn\dvacc $$
and
$$\cases{\chi_n\bra 0 =0 \qquad{\overline \chi}_m\bra 0 =0
 &in the Ramond sector\cr
\chi_{n+\hal}\bra 0 =0\qquad{\overline \chi}_{n+\hal}\bra 0 =0
   &in the Neveu-Schwarz sector\cr}\eqn\dvac$$
for $n\geq0$ and $m\geq 1$ and
where
$$\eqalign{&\varphi=(\,\beta,\psi^+,\,b,\,\beta_+)\cr
&\chi=(\psi^{(0)},\psi^{(-)},\psi^{(-\hal)},\gamma^+,\,{\wt b}\>) \cr}
\qquad\qquad
\eqalign{&{\overline \varphi}=(\gamma,\,\psi,\,c,\,\gamma_+)\cr
&{\overline \chi}=(\psi^{(+)},\psi^{(\hal)},\gamma_{z+},\,{\wt c}\>).\cr}$$
 Due to the  presence of the c-number function $\zeta$ in $\qb$,
the physical vacuum ${\bra 0}_{\ssr {phys}}$ is
$${\bra 0}_{\ssr {phys}}=\cases{e^{-{\wh k}\zeta^2_0\gamma_0}\,e^{-\zeta_0
\psi_0^{(\hal)}}\>\bra 0&\quad for the Ramond sector\cr
e^{-{\wh k}\zeta^2_{\hal}\gamma_{-1}}\,e^{-\zeta_{\hal}\psi^{(\hal)}_{-\hal}}\>
\bra 0&\quad for the Neveu-Schwarz sector\cr}$$
The multiplied factors in the physical vacuum of the Neveu-Schwarz sector
involve the nonzero modes $\psi^{(\hal)}_{-\hal}$ and $\gamma _{-1}$
which change the energy of the physical vacuum.  The effect of these
factors on the super stress-energy tensor will be discussed in the next
section.  Taking into account the nontrivial physical vacuum,  the explicit
form of the projection ${\cal P}$ is
$$\eqalign{ {\cal P}=\sum^{\infty}_{i=0} \sum^{\infty}_{j=0}
\sum_{0<n_1 <\cdots < n_i} \sum_{(m_1, \cdots, m_j)}&
{1\over j !}(\psi_{\ssr {phys}})_{-n_1-\kappa}\cdots
(\psi_{\ssr {phys}})_{-n_i-\kappa}
\;\phi_{-m_1}\cdots\phi_{-m_j} {\bra 0}_{\ssr {phys}}\cr
&\qquad\qquad \times {}_{\ssr {phys}}\!\!<0\,\vert\>\phi_{m_j}\dots\phi_{m_1}\;
(\psi_{\ssr {phys}})_{n_i+\kappa}\cdots (\psi_{\ssr {phys}})_{n_1+\kappa}\cr}
$$
and so it is annihilated by $\qb$.

 	In the case of the SL(2,R), one directly use the same argument
on the physical subspace as in gauge theories because its BRST charge consists
only of bilinear terms in the mode variables [\BOC].  On the other hand,
our BRST charge $\qb$ contains higher-order terms  up to biquadratic ones,
as well as bilinear terms.  We will use the following device,
which has been designed in [\BRST] to prove the no-ghost theorem
in  string theory based on the BRST formalism.
 Introducing a parameter $t$, a $t$-dependent BRST charge $\qb^t$ is
constructed from $\qb$ as
$$\qb(t)= A+t\,B+t^2\,C \eqn\qbt $$
where the operators $A$, $B$ and $C$ are
$$ \eqalign{A&\equiv\qb^{(0)}\equiv\oint dz\>\lbrace c_z
(\beta+{\textstyle {\hal}}
\zeta^2)+\gamma_+(-{\textstyle {\hal}}\zeta^2\psi-\psi^+
-{\textstyle{\sqrt{2\over{\wh k}}}}\>
\psi^{(0)}\zeta\>)+\gamma_{z+}\psi^{(-)} \cr
&\qquad\qquad\qquad\qquad\qquad\qquad\qquad\qquad
+{\wt c}_z(\psi^{(-\hal)}-{\textstyle{\sqrt{2{\wh k}}}}\>\zeta\>)\rbrace
\cr
B&\equiv\oint dz \> \bigl[ \>\qb^{(0)},{\textstyle{1\over{\wh k}}}
\>\lbrace 2c_z\psi^{(0)}\beta^+
-{\textstyle {1\over4}}c_z(\psi^{(-\hal)}
+{\textstyle{\sqrt{2{\wh k}}}}\>\zeta\>){\wt b}
-\gamma_+\psi^{(0)}{\wt b}-\gamma_+\psi^{(\hal)}\beta^+\rbrace \cr
&\qquad\qquad\qquad\qquad\qquad\qquad\qquad\qquad
      -\gamma_+\psi b\>\bigr]
 \cr
C&\equiv\oint dz \>{\textstyle{\sqrt{2\over{\wh k}}}}
     \>\zeta\>(\gamma_+c_z\beta^+
  +{\textstyle \hal}\gamma_+^2{\wt b} ) \cr } \eqn\abc $$
The operator $A(=\qb^{t=0})$ has the same bilinear form as the asymptotic one
of
the BRST charge in the ordinary gauge theories [\KOS], thus being refferred
as $\qb^{(0)}$. The $t$-dependent BRST charge $\qb^t$ is nilpotent
for any value of $t$, because the operators
$A$, $B$ and $C$ fortunately satisfy the following algebraic relations
$$\eqalign{&A^2=C^2=0\cr
&\{A, B\}=\{B,C\}=0\cr
&B^2+\{A,C\}=0.\cr}\eqn\relabc $$
When $t=1$, $\qb^{t=1}$ agrees with the original BRST charge $\qb$.
The definition of a $t$-dependent physical space $\hp (t)$ is obtained
 by replacing $\qb$ in the definition \dphys\ with the $t$-dependent $\qb^t$.
  Physical fields (anti)commutting with $\qb^t$ are
$$\eqalign{&\phi^t=\phi \cr
& \psi^t_{\ssr {phys}}=i{\cal N}(t)\{(t\beta\psi+\psi^+)
+(t-1){\textstyle{\hal}}\zeta^2
\psi\}\cr}$$
with  the normalization factor
${\cal N}(t)=\{2(\beta+(t-1)\hal\zeta^2)\}^{-{1\over2}}$.  A $t$-dependent
operator
${\cal P}^t$ is a projection onto the $t$-dependent space
${\cal H}(\psi^t, \psi^t_{\ssr {phys}})$, thus being annihilated by $\qb^t$.

  The $t$-dependent theory as defined above will become the device
for our proof of the general form of physical states. It will be proved
in the $t$-dependent theory that any state $\bra {{\Phi}^t}$ annihilated by
$\qb ^t$ has the form
$$ \bra{\Phi ^t} ={\cal P}^t\,\bra{\Phi^t}+\qb^t\bra{\ast,t}, \eqn\gft $$
using mathematical induction with respect to the power of $t$.
In our induction, all $t$-dependent states and operators are expanded in $t$
as $\bra{\Phi^t\!\!}=\sum^\infty_{n=0}\bra{\Phi^t_n\!\!}t^n$
and ${\cal O}^t=\sum^\infty_{n=0}{\cal O}^t_n\, t^n$, assuming their
nonnegative expansions.  A starting point in the induction is the $t=0$
theory to which the techniques developed for the unitarity proof in gauge
 theories [\KOS] can be applied. Since the operators $A$, $B$ and $C$
satisfy the same algebra \relabc\ as in string theory, the essential part
of our proof for \gft\ has been described in [\BRST].
In the following, we will give only a sketch for our proof
and recommend to read them for the details.

In the $t=0$ theory,  one can easily find out two physical fields and
four Kugo-Ojima (KO) quartets
$$\eqalign{&\phi^{t=0}=\phi\cr
&(\>\gamma,\>\beta+{\textstyle \hal}\zeta^2, \>c_z,\>b\>)\cr
&(\>\psi^{(+)},\>\psi^{(-)},\>\gamma_{z+},\>\beta^+\;)\cr}\qquad
\eqalign{& \psi_{\ssr {phys}}^{t=0}
         ={\textstyle{i\over\sqrt{-\zeta^2}}}({\textstyle\hal}
    \zeta^2\psi-\psi^+)\cr
&(\>\Psi^{(0)}_-, \>\Psi^{(0)}_+,\>\gamma_+,\>\beta_+\>)\cr
&(\>\psi^{(\hal)}, \>\psi^{(-\hal)}-{\textstyle{\sk}}\zeta,
           \>{\wt c}_z,\>{\wt b}\>)\cr}\eqn\pql $$
 where $\Psi^{(0)}_\pm \equiv \hal \zeta^2\psi+\psi^+
             \pm{\ss{\sqrt{2\over{\wh k}}}}\psi^{(0)}\zeta$.
Repeating the same argument for those physical fields and KO quartets as in
gauge theories [\KOS], one can show that the general solution for
$\qb^0\bra{\Phi_0}=0$ is
$$ \bra{\Phi_0}={\cal P}^{t=0}\bra{\Phi_0}+\qb^{(0)}\bra{\ast,0}
\eqn\gfl$$
where ${\cal P}^{t=0}$ is a projection operator onto the subspace
${\cal H}(\phi^{t=0},\psi^{t=0}_{\ssr {phys}})$.
The next step is to show by the induction that  the nth term
$\bra{\Phi^t_n}$ in the expansion $\bra{\Phi^t}=\sum^\infty_{n=0}
\bra{\Phi^t_n}t^n$ has the form
$$\bra{\Phi^t_n}=\sum^n_{m=0}{\cal P}^t_{n-m}\bra{\Phi^t_m}+A\bra{\ast,n}
+B\bra{\ast,n-1}+C\bra{\ast,n-2}\eqn\pnt$$
where ${\cal P}^t=\sum^\infty_{n=0}{\cal P}^t_n t^n$
and $\bra{\ast}=\sum^\infty_{n=0}\bra{\ast,n}t^n$, understanding that
$\bra{\ast,n}=0$ for $n <0$.  For $n=0$,  the expression \pnt\ is satisfied
by $\Phi ^t_0=\Phi _0$ because it   agrees with \gfl.
The condition for physical state, $\qb^t\bra{\Phi^t}=0$,
leads to a relation
$$ A\bra{\Phi^t_n}+B\bra{\Phi^t_{n-1}}+C\bra{\Phi^t_{n-2}}=0 \eqn\abcpn$$
  The statement \pnt\
for $n=N$ can be proved assuming one for $n\leq N-1$ and using \abcpn\ and
 the following relation
$$A\sum^n_{m=0}{\cal P}^t_{n-m}\bra{\Phi^t_m}
  +B\sum ^{n-1}_{m=0}{\cal P}^t_{n-1-m}\bra{\Phi^t_m}
    +C\sum ^{n-2}_{m=0}{\cal P}^t_{n-2-m}\bra{\Phi^t_m}=0 $$
derived from $\qb^t{\cal P}^t\bra{\Phi^t}=0$.  Multiplying the expressions
\pnt\ by $t^n$ and summing up them over $n$, we finally arrive at the
general form \gft\ of physical state in the $t$-dependent theory
to be proved.  The general form \physp\ of physical state
is obtained by setting $t=1$ in the expression \gft\ and implies that
the physical subspace $\hp$ is made up only of the scalar field $\phi$
and the fermion $\psi _{\ssr {phys}}$ ;
$\hp={\cal H}(\phi, \psi_{\ssr {phys}})$.

\chapter{Physical Superconformal Algebra}

   In the last section we have suggested from the comparison of their
anomalies $c^{\ssr {total}}$ and $c^{\ssr {SCFT}}$
that the subspace ${\cal H}_{\ssr {phys}}$
is the super Coulomb gas with background charge $\alpha _0(k)$.  To see it,
we will show that the total supercurrent $G^{\ssr {total}}$ becomes one of
the super Coulomb gas [\SCF] in the physical space,
$$ G^{\ssr {SCFT}}(z)=i\partial\phi\,\psi_{\ssr {phys}}
       +2\alpha_0(k)\partial\psi_{\ssr {phys}}.
  \eqn\scft $$
The supercurrent $G^{\ssr {total}}$, hence, plays a role of supersymmetry
generators both in the original space of the SKMA and ghost systems and the
physical one ${\cal H}_{\ssr {phys}}$.  Given the supercurrent,
the corresponding stress-energy tensor $T^{\ssr{SCFT}}$ is easily derived
from the operator product of two supercurrents,
$$T^{\ssr{SCFT}}=-{\textstyle\hal}(\partial\phi)^2
  +i\alpha_0(k)\partial^2\phi-{\textstyle\hal}\psi_{\ssr{phys}}\partial
                  \psi_{\ssr{phys}}. \eqn\scftt $$
Both of the operators $G^{\ssr{SCFT}}$ and $T^{\ssr{SCFT}}$ generate a
superconformal algebra in the physical space.

   The analytic properties of the fermions $\psi ^{(\ssr A)}$ determine ones
of the function $\zeta(z)$ through the constraint $\psi^{(-\hal)}-\sk\zeta=0$
so that our simple choices of $\zeta(z)$  are the constant $\zeta_0$
for the Ramond sector  and $\zeta_{\hal}z^{-\hal}$ for the Neveu-Schwarz
sector.  These analytic properties are carried into the fermion
$\psi_{\ssr{phys}}$ in the physical space by the factor $\sqrt{\beta}$
because of its definition \sqrb\ in terms of $\zeta(z)$. The Ramond and
Neveu-Schwarz sectors in the physical space, thus,  correspond to ones
of the SKMA.

The operator product of the supercurrent $G^{\ssr {total}}$ and
the scalar field $\phi$ gives its superpartner, which would be expected
to be the fermion $\psi_{\ssr {phys}}$,
$$ G^{\ssr total}(z)\,\phi(w)\sim{1\over z-w}\,i
{\textstyle{\sqrt{2\over{\wh k}}}}\,
(\>\psi^{(-)}\gamma
 +{\textstyle{\hal}}\psi^{(-\hal)}\psi
  -\psi^{(0)}\>) $$
Since it anticommutes with $\qb$, the operator in the right-hand side
of the above operator product can be written in the general form \physo\
of physical operator.  Its genuine part is really a $\phi$'s superpartner
in the space ${\cal H}_{\ssr{phys}}$ and it must be made up only of $\phi$
and $\psi_{\ssr{phys}}$ because of ${\cal H}_{\ssr{phys}}
={\cal H}(\phi,\psi_{\ssr{phys}})$ as we have shown in the last section.
 The physical operator can be rewritten in succession into a series of BRST
equivalent ones,
using the BRST charge $\qb$, to reach its desired form as described above ;
for example of such a process,
$$ \eqalign{ \psi^{(-)}\gamma
 +{\textstyle{1\over 2}}\psi^{(-\hal)}\psi
  -\psi^{(0)}={\wh k}\,\zeta\psi-\psi^{(0)}& -{\textstyle{\hal}}\gamma_+
{\wt b}-c_z\beta^+ \cr
 &\qquad+[\>\qb, \beta^+\gamma+{\wt b}\psi\>].\cr} $$
As a result of repeating this type of calculation,  the following identity
is obtained,
$${\textstyle{\sqrt{2\over{\wh k}}}}(\,{\wh k}\zeta\,\psi
-\psi^{(0)}-{\textstyle{\hal}}\gamma_+{\wt b}-c_z\beta^+\>)
  ={\textstyle{i\over\sqrt{2\beta}}}\>(\>\beta\psi+\psi^+)+[\>\qb,\ast\ast\>]
\eqn\idqb $$
 where  the definition \sqrb\ of $\sqrt{\beta}$ has been used
and $\ast\ast$ denotes some complicated operator.  Using those identities,
the above operator product is successfully rewritten as
$$ G^{\ssr {total}}(z)\,\phi(w)\sim{1\over z-w}\>
    \bigl\{\,i\psi_{\ssr {phys}}(w)+ [\>\qb, \ast\ast\ast]\bigr\}.\eqn\gtphi $$
 Using the operator product \gtphi\ and the superconformal algebra of
$ G^{\ssr{total}}$ and $T^{\ssr{total}}$, one can show that
$G^{\ssr{total}}$ also transforms
$\psi_{\ssr{phys}}$ into $\phi$,
$$ G^{\ssr{total}}(z)\psi_{\ssr{phys}}(w)\sim {\phi\over (z-w)^2}
        +{\partial\phi\over(z-w)}+[\>\qb, \cdots\>]\eqn\gtpsi$$
with use of the fact that $G^{\ssr{total}}$ anticommutes with $\qb$.
Thus, the supercurrent $G^{\ssr{total}}$ is a supersymmetry generator in the
space ${\cal H}(\phi,\psi_{\ssr{phys}})$.

   Let us show that  the genuine part of the supercurrent $G^{\ssr{total}}$
 is really one of the super Coulomb gas \scft.
The similar but much longer calculation gives the following expression of
the supercurrent
$$\eqalign{G^{\ssr{total}}=(\>i\partial\phi+2\alpha_0(k)\partial\>)
{\textstyle{\sqrt{2\over{\wh k}}}}(\>{\wh k}\,\zeta\psi-\psi^{(0)}
   -&{\textstyle{\hal}}\gamma_+{\wt b}-c_z\beta^+)
-2{\textstyle{\sk}}\alpha_0(k)\partial\zeta\psi \cr
&+[\>\qb, \cdots\>].\cr}$$
 For the Ramond sector with the function $\zeta$ constant,
substituting the identity \idqb\ into the above expression,
 the supercurrent has the following form to be expected,
$$G^{\ssr{total}}=G^{\ssr {SCFT}}+[\>\qb,\ast]\eqn\gfpg$$
 On the other hand,  for the Neveu-Schwarz ( and  Ramond with $\zeta$
dependent of $z$ ) sector(s),  an additional term
$\partial\zeta\,\psi$, which does not anticommute
with $\qb$, is involved in the  expression
 $$ G^{\ssr{total}}=G^{\ssr {SCFT}}+[\>\qb,\ast]
                       -2{\textstyle{\sk}}\,\alpha_0(k)\partial\zeta\psi.$$
Defining a new supercurrent ${\wt G}^{\ssr{total}}$
as
$${\wt G}^{\ssr{total}}\equiv G^{\ssr{total}}
                   +2{\textstyle{\sk}}\,\alpha_0(k)\partial\zeta\psi,
\eqn\newg $$
 ${\wt G}^{\ssr{total}}$ anticommutes with $\qb$ and has the same form
as \gfpg.
The corresponding stress-energy tensor ${\wt T}^{\ssr{total}}$ is
$${\wt T}^{\ssr{total}}\equiv T^{\ssr{total}}
   +\partial\zeta\>(\>2\gamma\psi^{(-)}\psi-2\psi^{(0)}\psi
           -\psi^{(-\hal)}\gamma-\psi^{(\hal)}\>). \eqn\newt $$
These BRST-invariant supercurrent and stress-energy tensor
${\wt G}^{\ssr{total}}$ and ${\wt T}^{\ssr{total}}$ are ones for
the Neveu-Schwarz (and Ramond with  $\zeta$ dependent of $z$) sector(s).
For our simple choice $\zeta(z)=\zeta_{\hal}\>z^{-\hal}$,
although $L^{\ssr{total}}_0\>\bra{0}_{\ssr{phys}}=(\>\hal\zeta_{\hal}
\psi^{(\hal)}_{-1}-{\wh k}\zeta_{\hal}^2\,\gamma_{-1}\>)
\bra{0}_{\ssr{phys}}$,  the energy of the physical vacuum with respect to ${\wt
T}^{\ssr{total}}$
is zero due to the added term of $\partial\zeta$ in \newt\ ;
${\wt L}^{\ssr{total}}_0\bra{0}_{\ssr{phys}}=0$.
Note that in the SL(2,R) case  one must add an $a$-dependent term to
the stress-energy tensor to get a BRST-invariant one when a constraint
 $J^{(-)}(z)-a(z)=0$ with a $z$-dependent $a(z)$ is imposed.

\chapter{Conclusion and Discussions}

 Taking the super Kac-Moody algebra of OSp(1,2), we have given the
constraints on superspace currents, which allow us to formulate
the Hamiltonian reduction in the supersymmetric way,
to reduce the SKMA into the superconformal algebra.
Constructing the BRST charge from those constraints in the standard way,
 the familiar BRST formalism has been applied to the above Hamiltonian
reduction.
With help of the free-field realization of the OSp(1,2) KMA,
we have analyzed the structure of the physical subspace and the relation of
superconformal algebras both in the SKMA and the physical subspace to show
that the subspace is really a space of the super Coulomb gas. This result
suggests that the hidden symmetry of superconformal field theory is the
OSp(1,2) super Kac-Moody algebra, because the whole space of the
SKMA is reduced into one of SCFT through the Hamiltonian reduction.

Finally we will give a few comments. In the present paper we have focused
on the relation between two spaces of free fields realizing SCFT and the
OSp(1,2) SKMA.  Although each of the free field realizations is
highly reducible, One can extract an irreducible representation
of the algebra from it
by using the screening charge as pointed out by Felder [\F].
Two screening operators ${\bf{\cal J}}^{\pm}_{\ssr{OSp}}$
of the OSp(1,2) free-field realization  are expressed in the superspace
$$\eqalign{{\bf{\cal J}}^-_{\ssr{OSp}}(z,\theta)=& \zeta^{-1}\>\bigl\{\>
{\textstyle{1\over\sqrt{{\wh k}}}}\>
(\>\psi^{(-)}\psi-{\textstyle{1\over2\sqrt{2}}}\>\psi^{(-\hal)}\>)
  +\theta\>(\>\psi^++\beta\psi\>)\bigr\}\>e^{i\alpha_-\phi}\cr
   {\bf{\cal J}}^+_{\ssr{OSp}}(z,\theta)=&\zeta^{-1}\>\bigl\{
\>{\textstyle{1\over\sqrt{{\wh k}}}}\>
\beta^{-k-3}\>(\>(k+2)\psi^{(-)}\psi^++(k+1)\psi^{(-)}\psi\beta
+{\textstyle{1\over2\sqrt{2}}}\>\psi^{(-\hal)}\beta\>)\cr
&\qquad\qquad\qquad\qquad\qquad\qquad\qquad\qquad+\theta\>
\beta^{-k-2}(\>\beta\psi+\psi^+\>)\bigr\}\>e^{i\alpha_+\phi}\cr}
\eqn\scrosp $$
On the other hand, ones of the super Coulomb gas are [\SCF]
$${\bf{\cal J}}^{\pm}_{\ssr{SCFT}}(z,\theta)=(\>1+\theta\psi_{\ssr{phys}}\>)
 e^{i\alpha_{\pm}\phi} \eqn\scgsc $$
These two sets of screening operators are BRST-equivalent to each other,
  $${\bf{\cal J}}^{\pm}_{\ssr{OSp}}(z,\theta)=
              {\bf{\cal J}}^{\pm}_{\ssr{SCFT}}(z,\theta)+[\>\qb,\ast\>].
 \eqn\relscr$$
The presence of the factor $\zeta^{-1}$ in the expression \scrosp\ has
allowed the screening operators ${\bf{\cal J}}^{\pm}_{\ssr{OSp}}$ to
satisfy the BRST-equivalent relation \relscr. Using the BRST-equivalent
relation \relscr, one can show in the same way as in [\BOC] that any
irreducible representation of the SKMA is reduced into the corresponding
one of SCFT via the Hamiltonian reduction. \break\hfill
\indent  The present construction can be extended to the case of OSp($N$,2)
$( N\geq2 )$ SKMA. In $N=2$ case the SKMA generates not only a N=1
superconformal algebra but also a N=2 one [\KS]. Thus one can expect that
the OSp(2,2) SKMA is reduced into the N=2 SCA in a manifestly N=2
supersymmetric way.

\vfill\supereject
\refout
\vfill\supereject
\bye
\end